\documentclass[twocolumn]{aastex631}
\usepackage{amsmath}
\usepackage{graphicx}
\usepackage[normalem]{ulem}
\usepackage{txfonts}
\usepackage{hyperref}

\usepackage{mathtools}
\usepackage{xcolor}

\shorttitle{3D orbit of AC\,Her}
\shortauthors{Anugu et al.}

\graphicspath{{./}{Figures_new/}}

\begin{document}

\title{Three-dimensional orbit of AC\,Her determined:  Binary-induced truncation cannot explain the large cavity in this post-AGB transition disk}

\correspondingauthor{Narsireddy Anugu}
\email{nanugu@gsu.edu}

\author[0000-0002-2208-6541]{Narsireddy Anugu}
\affiliation{The CHARA Array of Georgia State University, Mount Wilson Observatory, Mount Wilson, CA 91023, USA}
\affiliation{Department of Astronomy and Steward Observatory, University of Arizona, 933 N. Cherry Avenue, Tucson, AZ 85721-0065, USA}

\author[0000-0002-9491-393X]{Jacques Kluska}
\affiliation{Institute of Astronomy, KU Leuven, Celestijnenlaan 200D, 3001 Leuven, Belgium}

\author[0000-0002-3003-3183]{Tyler Gardner}
\affiliation{School of Physics and Astronomy, University of Exeter, Exeter, Stocker Road, EX4 4QL, UK}

\author[0000-0002-3380-3307]{John D. Monnier}
\affiliation{University of Michigan, Ann Arbor, MI 48109, USA}

\author[0000-0001-5158-9327]{Hans Van Winckel}
\affiliation{Institute of Astronomy, KU Leuven, Celestijnenlaan 200D, 3001 Leuven, Belgium}

\author[0000-0001-5415-9189]{Gail H. Schaefer}
\affiliation{CHARA Array, Georgia State University, Atlanta, GA 30302, USA}

\author[0000-0001-6017-8773]{Stefan Kraus}
\affiliation{School of Physics and Astronomy, University of Exeter, Exeter, Stocker Road, EX4 4QL, UK}

\author[0000-0002-0493-4674]{Jean-Baptiste Le~Bouquin}
\affiliation{Institut de Planetologie et d’Astrophysique de Grenoble, Grenoble 38058, France}

\author[0000-0002-2314-7289]{Steve Ertel}
\affiliation{Department of Astronomy and Steward Observatory, University of Arizona, 933 N. Cherry Avenue, Tucson, AZ 85721-0065, USA}
\affiliation{Large Binocular Telescope Observatory, University of Arizona, 933 N. Cherry Avenue, Tucson, AZ 85721-0065, USA}

\author[0000-0003-2125-0183]{Antoine M\'erand}
\affiliation{European Southern Observatory,  85748 Garching, Munich, Germany}

\author[0000-0002-2488-7123]{Robert Klement}
\affiliation{CHARA Array, Georgia State University, Atlanta, GA 30302, USA}

\author[0000-0001-9764-2357]{Claire L Davies}
\affiliation{School of Physics and Astronomy, University of Exeter, Exeter, Stocker Road, EX4 4QL, UK}

\author[0000-0002-1575-4310]{Jacob Ennis}
\affiliation{University of Michigan, Ann Arbor, MI 48109, USA}

\author[0000-0001-8837-7045]{Aaron Labdon}
\affiliation{European Southern Observatory, Casilla 19001, Santiago 19, Chile}

\author[0000-0001-9745-5834]{Cyprien Lanthermann}
\affiliation{CHARA Array, Georgia State University, Atlanta, GA 30302, USA}

\author[0000-0001-5980-0246]{Benjamin R. Setterholm}
\affiliation{University of Michigan, Ann Arbor, MI 48109, USA}

\author[0000-0002-0114-7915]{Theo ten Brummelaar}
\affiliation{CHARA Array, Georgia State University, Atlanta, GA 30302, USA}

\author[0000-0001-9539-0942]{Akke Corporaal} 
\affiliation{Institute of Astronomy, KU Leuven, Celestijnenlaan 200D, 3001 Leuven, Belgium}

\author[0000-0003-0242-0044]{Laurence Sabin}
\affiliation{Instituto de Astronom\'ia, Universidad Nacional Aut\'onoma de M\'exico, AP 106, Ensenada 22800, B.C., M\'exico}

\author[0000-0002-2488-7123]{Jayadev Rajagopal}
\affiliation{NSF’s National Optical-Infrared Astronomy Research Laboratory, 950 N. Cherry Ave., Tucson, AZ 85719, USA}

\begin{abstract}
Some evolved binaries, namely post-asymptotic giant branch binaries, are surrounded by stable and massive circumbinary disks similar to protoplanetary disks found around young stars. Around 10\% of these disks are transition disks: they have a large inner cavity in the dust.
Previous interferometric measurements and modeling have ruled out the cavity being formed by dust sublimation and  suggested that the cavity is due to a massive circumbinary planet that traps the dust in the disk and produces the observed depletion of refractory elements on the surface of the post-AGB star.
In this study, we test alternative scenario in which the large cavity could be due to dynamical truncation from the inner binary.
We performed near-infrared interferometric observations with the CHARA Array on the archetype of such a transition disk around a post-AGB binary: AC\,Her.
We detect the companion at ten epochs over 4 years and determine the 3-dimensional orbit using these astrometric measurements in combination with the radial velocity time series. 
This is the first astrometric orbit constructed for a post-AGB binary system. 
We derive the best-fit orbit with a semi-major axis $2.01 \pm 0.01$\,mas  ($2.83\pm0.08$\,au), inclination $(142.9 \pm 1.1)^\circ$ and longitude of the ascending node $(155.1 \pm 1.8)^\circ$.
We find that the theoretical dynamical truncation and dust sublimation radius are at least $\sim3\times$  smaller than the observed inner disk radius ($\sim21.5$\,mas or 30\,au).
This strengthens the hypothesis that the origin of such a cavity is due to the presence of a circumbinary planet.
\end{abstract}

\keywords{Post-asymptotic giant branch stars(2121) -- Long baseline interferometry(932) -- Binary stars(154) -- techniques: high angular resolution}

\section{Introduction} \label{sec:intro}

It is well established that planet formation takes place in protoplanetary disks (PPDs) around young stars.
These PPDs are extensively studied to trace the initial conditions and mechanisms of planet formation.
Spatially resolved observations of these disks revealed the presence of a wide diversity of structures such as gaps, spirals, or warps \citep[e.g.][]{Calvet2002ApJ...568.1008C,  Johansen2007Natur.448.1022J, Andrews2011, Pinilla2012A&A...545A..81P,  Sheehan2018ApJ...857...18S, Rich2022AJ....164..109R}.
These structures  can be linked to planet formation, whether they are caused by already-formed planets or will trigger planet-formation mechanisms is not yet understood.
The first evidence for gaps and inner cavities in the dust disks around PPDs was deduced from the spectral energy distribution (SED) \citep[e.g.][]{Strom1989AJ.....97.1451S}, where a lack of near-infrared excess was observed, while  for a full continuous disk hot, thermal dust emission is expected to be observed at these wavelengths. 
Various interpretations for the origin of such extended disk cavities have been proposed: the most likely being the presence of a giant planet or a companion or the action of disk dissipation through photo-evaporation \citep[e.g.][]{Espaillat2014prpl.conf..497E}.
Disks around young stellar objects with such cavities are called transition disks \citep{vanderMarel2023EPJP..138..225V}.

Interestingly, the presence of disks which large cavities were also detected around evolved stars, namely post-Asymptotic Giang Branch (post-AGB) binaries \citep{Hillen2015, Kluska2022, Corporaal2023arXiv230412028C}.
%
%
Post-AGB stars are luminous objects of low and intermediate initial masses ($0.8–8M_{\odot}$) but in their almost final stages of stellar evolution.
A post-AGB star has lost its envelope in the AGB phase and is now contracting to become a white dwarf, crossing the Hertzsprung-Russel diagram at an almost constant high luminosity. 
The companion star is likely a main-sequence star reported from spectroscopic observations \citep{Oomen2018}. 
Post-AGB stars with a main sequence companion are surrounded by massive dusty and gaseous disks as  indicated by a strong observational link between a disk-like infrared excess in the SED and a radial velocity detection of a companion \citep[see review, ][and references therein]{VanWinckel2018arXiv180900871V}.
These disks are strikingly similar to PPDs. 
They have similar masses \citep[e.g.,][]{Bujarrabal2015A&A...575L...7B, Sahai2011ApJ...739L...3S}, are stable \citep[i.e., in Keplerian rotation, e.g.,][]{Bujarrabal2013A&A...557L..11B, Bujarrabal2015A&A...575L...7B, Bujarrabal2018A&A...614A..58B, GallardoCava2021A&A...648A..93G}, and can be modeled with radiative transfer models of PPDs \citep{Hillen2015,Kluska2018}.
The disk's inner rim  can be resolved with infrared interferometry and is located at the dust sublimation radius for most of the targets \citep{Hillen2016A&A...588L...1H, Hillen2017A&A...599A..41H, Kluska2019,  Kluska2020A&A...642A.152K, Corporaal2021A&A...650L..13C}.

  Recently, the infrared excesses of all identified post-AGB binaries in the Galaxy (85 objects) was studied to characterize these disks \citep{Kluska2022}. While most of the infrared excesses are compatible with full disks \citep[categories 0 and 1 in][]{Kluska2022} around 10\% of the disks have a lack of near-infrared excess \citep[categories 2 and 3][]{Kluska2022}.
These disks are reminiscent of transition disks around  young stars, as the lack of near-infrared excess emission points to an inner rim at several times the dust sublimation radius.

Moreover, it was found that post-AGB stars surrounded by such transition disks are more depleted of refractory elements \citep{Kluska2022}.
This is a signature of a mechanism  trapping the dust in the disk, and letting the volatile elements be accreted onto the binary stars\footnote{The main-sequence companion is too faint compared to the primary post-AGB to be able to derive abundances, however, we assume both the primary and the companion accrete the depleted gas from the circumbinary disk.}.
A similar phenomenon is observed around Herbig stars, where the radiative nature of the stellar atmosphere makes it possible to relate the observed element abundances to the accretion history \citep{Kama2015A&A...582L..10K}. 
As the remaining envelope material is  low in mass in post-AGB stars, the accretion of depleted gas impacts much stronger on the surface abundances and depletion is much stronger than in Herbig stars \citep{Jermyn2018,Oomen2019}.
An efficient pressure bump in the disk, trapping the dust grains but letting the gas through, is expected to produce such depletion.
Thus, the main hypothesis to explain the large cavities are either  
 the inner binary which truncates the disk, or the presence of a third (planetary-mass) component in the system which truncates the disk and traps the dust in the disk \citep{Kluska2022}, while only the latter mechanism provides a scenario for depletion.

In this work, we study the post-AGB system AC\,Herculis \citep[AC\,Her, HD 170756, HIP 90697;][]{Hillen2015} that is a F4Ibp spectral-type and single-lined star spectroscopic binary with an orbital period of $1194\pm6$\,days \citep{VanWinckel1998A&A...336L..17V, Oomen2018}.
It is located at a distance of $1402 \pm 43$\,pc \citep[Gaia DR3][]{GaiaCollaboration2016A&A...595A...1G,GAIA2022}. We considered Gaia DR3 ``distance\_gspphot" distance for its lower uncertainty in comparison to other reported distance measurements, see Section~\ref{Sec:DynamicalMasses}. In what follows, we stress that our angular scales are mainly distance independent, while the translation to physical scales is.
AC\,Her is considered as the prime example of a transition disk around post-AGB binaries and is classified as category~2 in \citet{Kluska2022}.  AC\,Her is a RV\,Tauri pulsator and some basic properties are given in the Table~\ref{Table:ac_properties}.
Several tracers, including [Fe/H] of -1.5, reveal that the observed composition of the post-AGB star surface is depleted of refractory materials \citep{VanWinckel1998A&A...336L..17V} similarly in post-AGB transition disks \citep{Kluska2022}.

Observations with the Plateau de Bure Interferometer in $^{12}$CO~J~=~2-1 show that the circumbinary disk around AC\,Her is stable, i.e., in Keplerian rotation, with an outer radius of $\sim 1000$\,au in gas \citep{Bujarrabal2015A&A...575L...7B, GallardoCava2021A&A...648A..93G}.
Mid-infrared high angular observations with MIDI combined with a study using radiative transfer models of disks have shown that the inner disk cavity radius $21.5^{+5.0}_{-2.5}$ mas  ($30^{+7}_{-4}$\,au using the Gaia DR3  $1402 \pm 43$\,pc distance), which is around ten times larger than the theoretical dust sublimation radius \citep[1.5-5.0\,au from,][]{Hillen2015}.
The disk is inclined $(i=50\pm8)^\circ$ with a position angle of $PA=(305\pm10)^\circ$ \citep{Hillen2015}.

Here we investigate the origin of the large disk cavity in the AC\,Her's disk with high-angular resolution observations in the near-infrared with the CHARA Array \citep{tenBrummelaar2005,Gies2022SPIE12183E..03G}. 
The observation  and data reduction are described in Section~\ref{Sec:Observations}. 
The binary detection and detection limits, our best-fit orbital solution, and dynamical mass determinations are reported in Section~\ref{Sec:Results_Orbit}. 
The binary truncation and dust sublimation resulting in disk cavities are computed in the discussion Section~\ref{Sec:Discussion}. In the final Section, we summarize our findings and give conclusions.

\begin{deluxetable}{l l l l l l l l l }
\tablecaption{Properties of AC Her}
\tablewidth{0pt}
\tablehead{
\colhead{} & Parameter & Value & Ref } \label{Table:ac_properties} 
\startdata
Literature  & Spectral type & F4Ibp              & \\
            & $R_{\rm in}$ & $30^{+7}_{-4}$\,au              & 1\\
            &              & $21.5^{+5.0}_{-2.5}$\,mas       &  \\
            & $d$                     & $1402 \pm 43$\,pc    & 2\\
            & $L_*$                   & $2475~L_\odot$       & 3\\
            &                         & $10^4~L_\odot$       & 4\\
            & $T_\mathrm{eff}$        & $5255\pm125 - 5800\pm250$\,K & 1,6\\
Measured    & $a$                     & $2.83\pm0.08$\,au    & 7\\
            &                         & $2.01 \pm 0.01$\,mas & \\
            & $e$                     & $0.206\pm0.004$\,au  & 7\\
            & $q$                     & $0.52$               & 7\\
Derived     & $R_\mathrm{trunc}$      & $9.03\pm0.28$\,au    & 7\\
            &                         & $6.44\pm0.04$\,mas   &  \\
            & $R_\mathrm{subli}$ at $L_*=2475L_\odot$  & $3.2$\,au   & 7\\
            &                         & $2.3$\,mas           &   \\
            &$R_\mathrm{subli}$ at $L_*=10^4L_\odot$   & $9\pm1$\,au   & 7\\
            &                        & $6.4\pm0.5$\,mas            &   \\
\enddata
\tablecomments{~References 1. \citet{Hillen2015}, 2. \citet{GAIA2022}, 3. \citet{Bodi2019},  4. \citet{MillerBertolami2016}, 5. \citet{Kluska2019}, 6. \citet{Kluska2022} and 7-this work. The symbols, $R_{\rm in}$ is the radiative transfer model and observed interferometric data fitted inner disk cavity radius,  $d$ is the Gaia DR3 distance, $L_*$ is the post-AGB star luminosity, $T_\mathrm{subli}$ is the dust sublimation temperature, $Q_\mathrm{R}$ is the ratio of dust absorption efficiencies,  $q$ is the binary mass-ratio, $R_\mathrm{trunc}$ is the binary truncation radius,  $R_\mathrm{subli}$ is the dust sublimation radius. }
\end{deluxetable}

\section{Observations and data reduction} \label{Sec:Observations}

We used MIRC-X ~\citep{Anugu2018SPIE10701E..24A, Anugu2020AJ....160..158A,  Kraus2018SPIE10701E..23K}, a six telescope beam combiner instrument at the Center of High-Angular Resolution Astronomy (CHARA) Array, to obtain observations of AC\,Her in $H$-band (wavelengths $\lambda=1.4-1.72~\micron$).  
Between 2019 and 2022, we obtained observations at ten epochs.
The CHARA Array delivers high-angular resolutions in optical/near-infrared wavelengths using baselines from 30 to 330-m corresponding to an angular resolution of, e.g., $\lambda/{2B_{\rm max}}\sim0.5$~mas in H-band. 
The MIRC-X instrument measures 15 squared visibilities ($V^{2}$) and 20 closure phases ($T3PHI$) simultaneously.

The observations are obtained with low spectral resolutions of $\mathcal{R} = \lambda/\Delta\lambda \approx 50$, which provide eight spectral channels.
It results in a $\sim50$\,mas interferometric field-of-view that is enough to image the AC\,Her system \citep{Anugu2020AJ....160..158A}. 
MIRC-X provides precision orbits with astrometric measurements of close binaries down to $\sim10~\mu$as level \citep[e.g.][]{Gardner2021AJ....161...40G}. 
We used unresolved calibrator stars to calibrate the instrumental and atmospheric transfer functions. 
The details of the observational setup and instrument configurations with calibrators for each epoch are tabulated in Appendix~Table~\ref{Table:observations}. 
Our observations benefited from the recent installation of adaptive optics systems for the CHARA Array telescopes, which improved sensitivity by a magnitude \citep[e.g.][]{Che2013JAI.....240007C,tenBrummelaar2018SPIE10703E..04T,Anugu2020SPIE11446E..22A}.

The MIRC-X data are reduced with the standard \texttt{mircx pipeline}\footnote{\href{https://gitlab.chara.gsu.edu/lebouquj/mircx\_pipeline.git}{ https://gitlab.chara.gsu.edu/lebouquj/mircx\_pipeline.git}} version 1.3.5, which is publicly available \citep{Anugu2020AJ....160..158A}.  
To increase the signal-to-noise ratio (SNR) of $V^2$ and $T3PHI$, we reduced the data with an integration time equal to the atmospheric coherence time for each observation. We checked for any signs of binarity of the calibrators by calibrating calibrators against each other and searched for binarity signal in them using the \texttt{CANDID}~\citep{Gallenne2015} software. No signs of binarity were found among the calibrators.

\begin{figure*}[!htb]
\centering
\includegraphics[width=\textwidth]{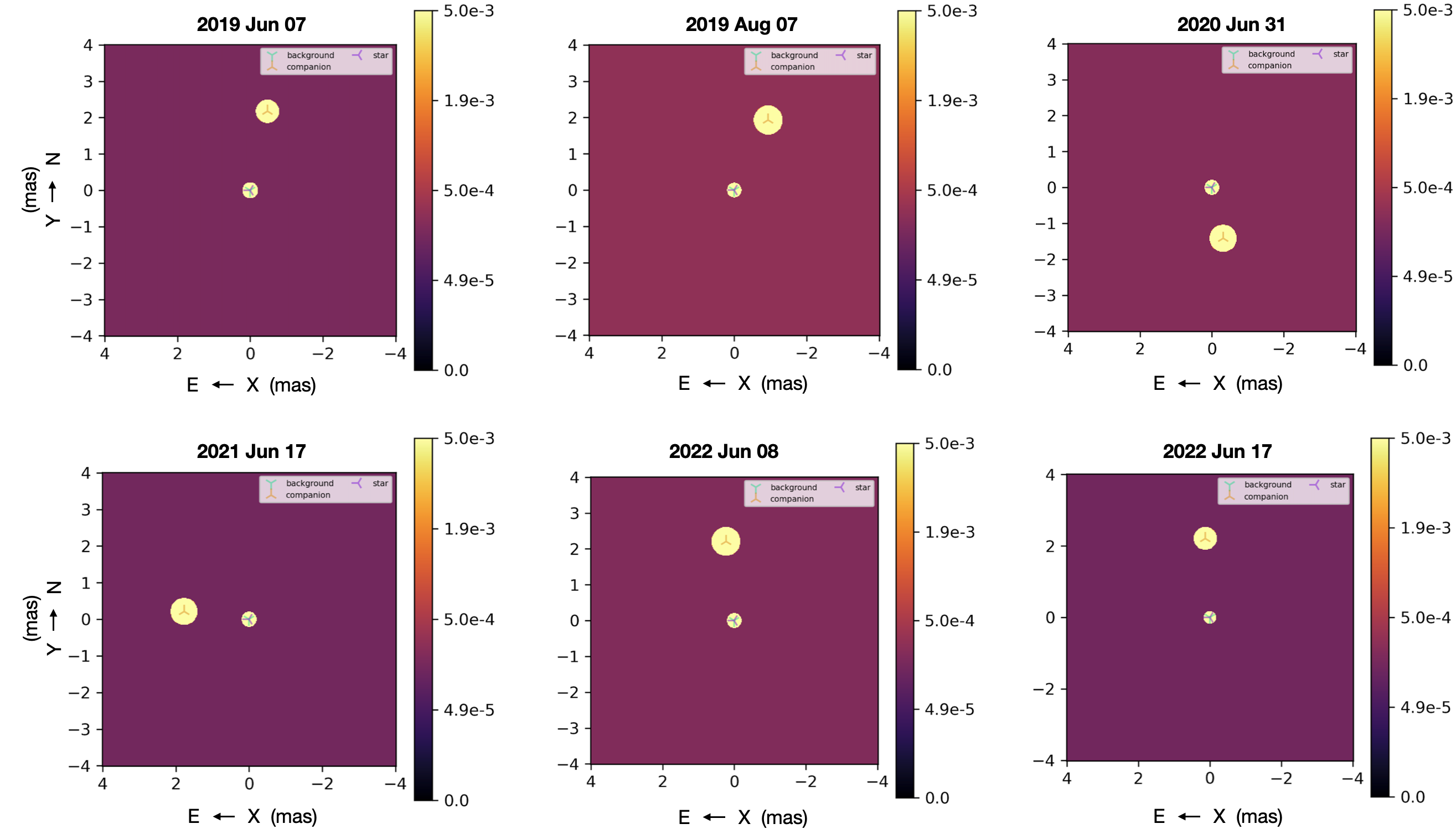}
\caption{The best-fit images constructed for various epochs spanning four observing years. The epoch is denoted on the top of each image window.  The images consist of the post-AGB star (smaller in size) and companion star (larger in size). The image is shown with a cut of $5\times10^{-3}$ of the surface brightness. North is up, and east is to the left. The field of view of this image is  $8 \times 8$ mas. }
\label{Figure:images}
\end{figure*}

\begin{figure*}[!htb]
\centering
\includegraphics[width=0.49\textwidth]{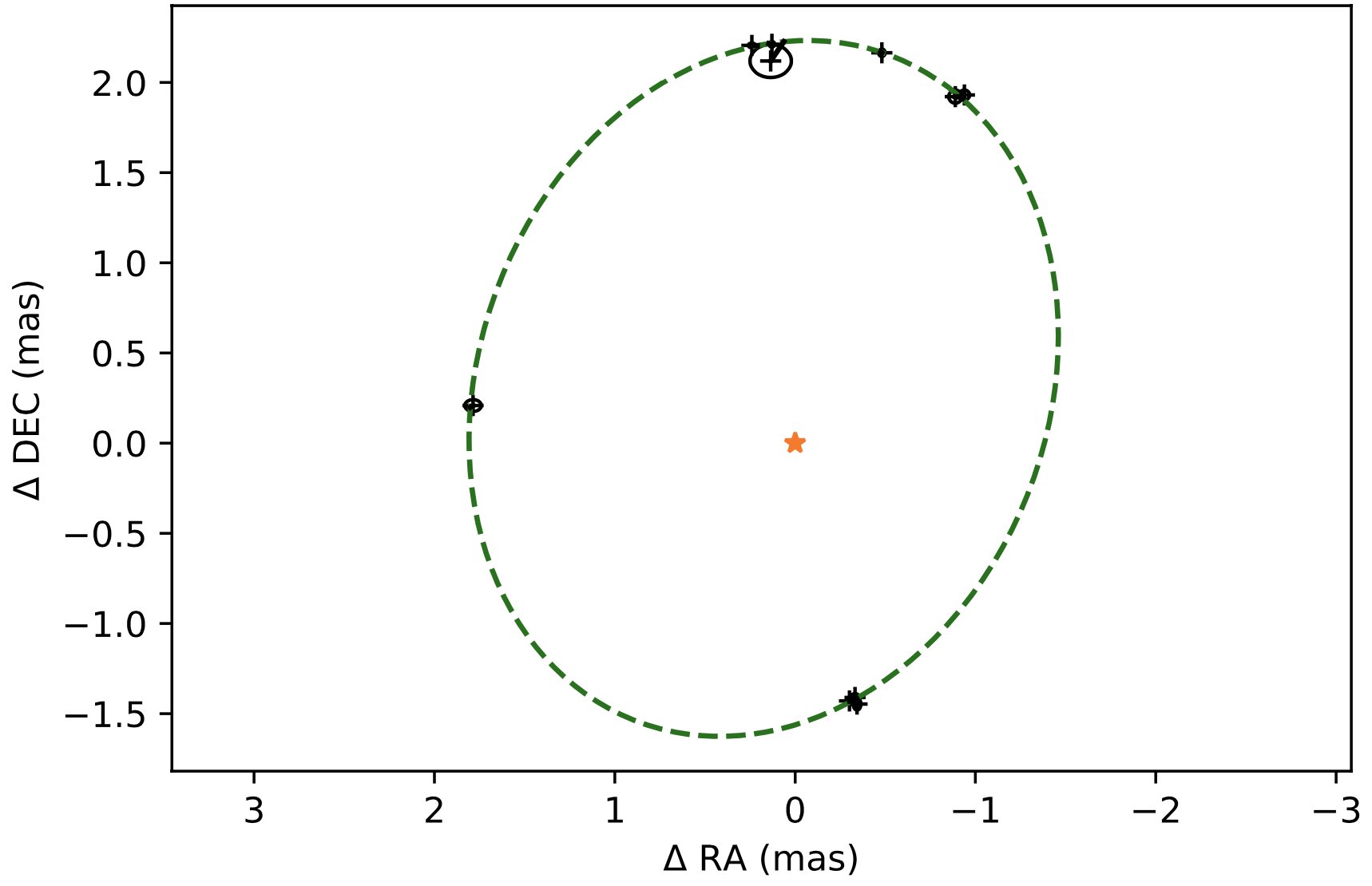}
\includegraphics[width=0.49\textwidth]{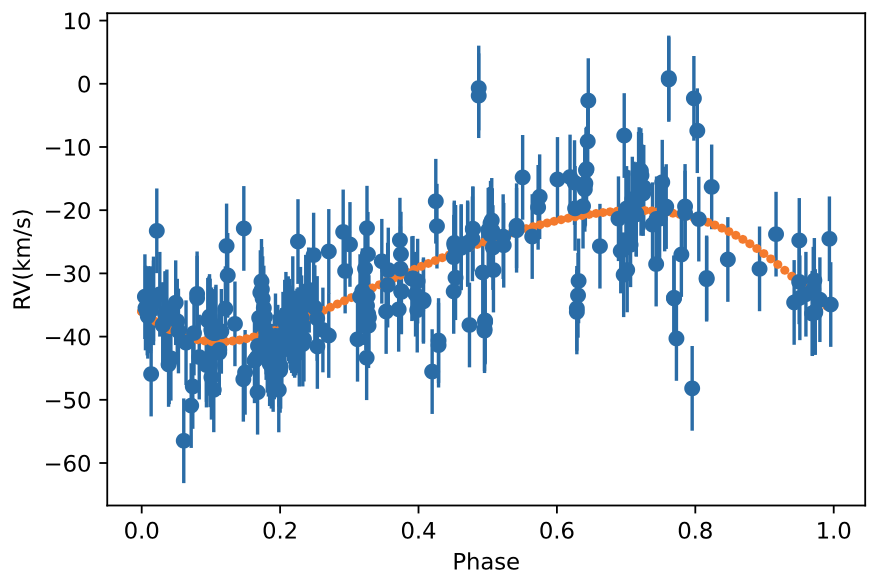}
\caption{The best-fit relative orbit of the companion around the primary post-AGB star. Left: Visual orbit of AC\,Her based on MIRC-X measurements at the CHARA Array.   The primary post-AGB is fixed at the (0,0) position and denoted with  orange ``$\star$" star symbol at the center. The astrometric  positions and  their uncertainties are denoted with   black ``+" and ellipse symbols. Right: Radial Velocity fit  based on the data from HERMES spectrograph at the 1.2-m Mercator telescope.  The orange color line is the best-fit to the data, which are filled circled points with error bars in blue color.} 
\label{Figure:binary_orbit}
\end{figure*}

\begin{deluxetable*}{l l l l l l l l l }
\tablecaption{Orbital elements of the AC Her system derived from combined interferometric astrometry and radial velocity data. 
}
\tablewidth{0pt}
\tablehead{
\colhead{Orbital element} & This work (Astrometry+RV) & Oomen et al. 2018 (RV) & Van Winckel et al. 1998 (RV)} \label{Table:orbit_fit} 
\startdata
Semi-major axis, $a$ (mas)         & $2.01 \pm 0.01$   & -                  & - \\
Inclination, $i$~($^\circ$)        & $142.9 \pm 1.1$   & -                  & -\\
$\Omega$($^\circ$)                 & $155.1 \pm 1.8$   & -                  &  -           \\
$\omega_1$($^\circ$)                 & $118.6 \pm 2.0$   & -                  & $114 \pm  12 $ \\
$T_0$ (MJD)                        & $59023.1 \pm 2.2$ & -                  &  $47129 \pm 35$ JD \\
eccentricity, $e$                  & $0.206 \pm 0.004$ & $0.0 \pm 0.05$     &  $0.12 \pm 0.02$\\
Orbital period, $P$ (days)         & $1187.7 \pm 0.7$  & $1188.9 \pm 1.2$   & $1194 \pm 6$ \\
$M_{\rm total}$  ($M_\odot$)      & $2.13 \pm 0.19$   & $0.75   \pm 0.03$  & $1.8 \pm 0.4$\\
$M_1$ ($M_\odot$)                  & $0.73 \pm 0.13$   & $0.6$ (fixed)      & $0.6 \pm 0.2$    \\
$M_2$ ($M_\odot$)                  & $1.40 \pm 0.12$   & $0.15 \pm 0.03$    & $1.2 \pm 0.2$\\
$K_1$ (km/s)                       & $10.5 \pm 0.5$    & $10.8 \pm 0.7$     & $12.7 \pm 0.3$  \\
$\gamma$ (km/s)                    & $-29.3 \pm 0.4$   & $-27.0 \pm 0.2$    & $-33$  \\
\enddata
\tablecomments{$P$ denotes the orbital period, $a$ the orbital semi-major axis, $e$ the eccentricity, $i$ the orbital inclination, $T_0$ the time of periastron passage,   $K_1$ the semi-amplitude in the radial velocity of the stellar components, and $\gamma$ the velocity of the system center-of-mass. $\omega_1$ is the longitude of the periastron, measured from the ascending node of the companion, and $\Omega$ gives the longitude of the ascending node (i.e., the node where the motion of the companion is directed away from the post-AGB).}
\end{deluxetable*}


\begin{deluxetable*}{l l l l l l l l l }\label{Table:ACHer_mass_estimates}
\tablecaption{AC\,Her binary system dynamical masses for all reported distances}
\tablewidth{0pt}
\tablehead{
\colhead{Source [Ref]} & Distance [pc] & $M_\mathrm{total}$ $[M_\odot]$  & $M_1 [M_\odot]$ & $M_2 [M_\odot]$ } 
\startdata
\citet{Hillen2015}                & $1600\pm 300$& $3.1\pm1.8$     & $1.3\pm1.1$   & $1.8\pm0.7$ \\
Gaia DR2 Geometric            [1] & $1230\pm 44$   & $1.44\pm0.16$ & $0.36\pm0.10$ & $1.08\pm0.10$ \\
Gaia DR3 Geometric            [2] & $1626 \pm 94$  & $3.3\pm0.6$   & $1.4\pm0.4$   & $1.88\pm0.25$ \\
Gaia DR3 Photogeometric       [2] & $1620 \pm 76$  & $3.2\pm0.5$   & $1.41\pm0.31$ & $1.86\pm0.21$ \\
Gaia DR3 ``distance\_gspphot" [3] & $1402 \pm 43$  & $2.11\pm0.21$ & $0.73\pm0.14$ & $1.40\pm0.12$ \\
\enddata
\tablecomments{[1] -- \citet{Bailer-Jones2018AJ....156...58B}; [2] -- \citet{Bailer-Jones2021AJ....161..147B}; [3] -- \citet{GAIA2022}; }
\end{deluxetable*}

\section{Orbital solution and dynamical masses of the system\label{Sec:Results_Orbit}}

The  reduced MIRC-X data (see Appendix~Figure~\ref{Figure:fitresiduals}) show a significant non-zero closure phase signal. In this section, we present geometric models that show that these phase signals can be attributed to the binary system. We derive the relative astrometry for the two stars and fit the astrometric orbit.

\subsection{Companion detection\label{Sec:BinarySearch}}

To reproduce the data, we used  Python based Parametric Modeling of Optical InteRferomEtric Data  \citep[\texttt{PMOIRED\footnote{\href{https://github.com/amerand/PMOIRED.git}{https://github.com/amerand/PMOIRED.git}}}, ][]{Merand2022SPIE12183E..1NM}. 
 This software allows implementing geometric models with a combination of basic building blocks such as uniform disks, rings and Gaussian intensity distributions and their emission characteristics to fit the measured interferometric data in Fourier space.
We include a geometric model to fit the data ($V^2$ and $T3PHI$), where our model consists of a binary and background flux, the latter to recover the overresolved visibilities at the short baselines.  
The model has thus 7 parameters: the flux ratio of the companion  (\texttt{"companion,f"}), the position  (\texttt{"X", "Y"}) of the companion relative to the primary post-AGB star, and the uniform diameters of the two stars  (\texttt{"star,ud"}, \texttt{"companion,ud"}). Finally, the spectral indices of the companion uniform disk  (\texttt{"companion,ind"}) and the background  (\texttt{"background,ind"}).
The total flux is fixed at 1. This means that the fluxes are fitted as ratios to the total flux.

The astrometric positions of the companion and the rest of the geometric model fitting parameters are tabulated in Appendix~Table\,\ref{Table:astrometric_positions} with the associated statistical errors.
Figure~\ref{Figure:images} shows the geometrical model images for different epochs.

We find the companion at 10 epochs with separations from 1.46\,mas to 2.22\,mas. 
The derived astrometric positions match with those found with the \texttt{CANDID\footnote{\href{https://github.com/amerand/CANDID.git}{https://github.com/amerand/CANDID.git}}} \citep{Gallenne2015} binary search software to within $<2\sigma$, standard deviation error. 
The flux ratio of the companion to the post-AGB varies between 4.1 and 5.9\% of the total flux. 
The variations in the flux are attributed to the pulsations of the post-AGB star \citep[period 75.46\,d, ][]{Samus2017}. 

The post-AGB primary  uniform disk diameter is not resolved, i.e., $<0.5$\,mas. On the other hand, 
the uniform disk  of the companion is resolved  and has a $\sim0.75$\,mas diameter size (see Appendix~Table\,\ref{Table:astrometric_positions}). We note that this $\sim0.75$\,mas size is too large ($\sim 1$\,au.) to be the photosphere of the main sequence companion. 
%
%
We infer that this resolved component around the companion is likely the accretion disk, which launches the jet found in AC\,Her \citep{Bollen2022arXiv220808752B}. 
 The study on the origin of the circum-companion disk and its connection to jets is out of the scope of this paper, however. A separate report is in preparation.

\subsection{Astrometric and radial velocity orbital fit\label{Sec:Orbit_fit}}
 We constructed a 3-dimensional astrometric orbit by combining our interferometric astrometric positions with the already published radial velocity data \citep[][publicly available on VizieR]{Oomen2018} from the HERMES spectrograph \citep{Raskin2011} on the 1.2-m Mercator telescope, La Palma.

 We employed the orbital fitting tools reported in \citet{Gardner2021AJ....161...40G} to fit a binary model simultaneously to the interferometric astrometric positions and radial velocity data. 

 The radial velocity data are polluted with a scatter with an amplitude of $\sim15$\,km\,s$^{-1}$ because of irregular pulsations of the post-AGB star \citep{Oomen2018}. 
 We scaled the astrometric uncertainties by $1.86\times$ and RV uncertainties to 6.7\,km/s uniformly 
 for all points so that each data set, RV and interferometric astrometry, contributed a fixed $\chi^2_{\rm red}=1$.
Figure\,\ref{Figure:binary_orbit} shows the best-fit orbit, and Table\,\ref{Table:orbit_fit} presents the best-fit seven Campbell orbital parameters. 
The interferometric data mostly dominate the orbital fit. We obtain a median residual to the binary astrometry fit of $29~\mu$as.

The best-fit orbit solution  shows the binary orbit is highly inclined $i=142.9 \pm 1.1^\circ$ and eccentric $e=0.206 \pm 0.004$ with semi-major axis $a=2.01 \pm 0.01$\,mas, i.e.\ $2.83 \pm 0.08$\,au at the Gaia DR3 distance of $1402 \pm 43$\,pc. Our measured orbital period $P=1187.7 \pm 0.6$\,days is within $1\sigma$ of a previously-published radial velocity orbit solution \citep{VanWinckel1998A&A...336L..17V, Oomen2018}.

\subsection{Dynamical masses of the system}\label{Sec:DynamicalMasses}

Using Kepler's third law ($P^2 = a^3/M_{\rm total}$), we obtain the system's total dynamic mass of $M_\mathrm{total}=2.13 \pm 0.19 M_\odot$. We also computed this total mass for all previously reported distances as tabulated in Table~\ref{Table:ACHer_mass_estimates}. This Gaia distance was obtained using a single-star astrometric fit. The larger uncertainties in Gaia measurements may be because of the poor astrometric fit of the data with larger RUWE $\sim 2.1$ (Renormalised Unit Weight Error – goodness-of-fit statistic).

We measure the model-independent individual masses by substituting the binary orbital period ($P$), eccentricity ($e$) and inclination ($i$), and  semi-amplitude in the radial velocity of the stellar components ($K_1$) in the binary mass function \citep[e.g.][]{Oomen2018}:

\begin{equation} \label{eqn:dynamical_mass}
    f(m) \equiv \frac{K_1^3 P\, \big(1 - e^2\big)^{3/2}}{2\pi G} = \frac{\left(M_2 \sin i\right)^3}{M_\mathrm{total}^2}
\end{equation}

From Eq.~\ref{eqn:dynamical_mass},  we get individual masses for the companion ($M_2=1.40 \pm 0.12 M_\odot$) and post-AGB star ($M_1 = M_\mathrm{total}-M_2 =  0.73 \pm 0.13 M_\odot$).

This mass estimate of a post-AGB star compares well with typical white dwarf masses.
Considering the mass of the companion ($M_2=1.40 \pm 0.12 M_\odot$), we corroborate that the companion is most likely a main sequence star.

\section{Discussion on the origin of the disk cavity \label{Sec:Discussion} }

\begin{figure*}[!t]
\centering
\includegraphics[width=0.99\textwidth]{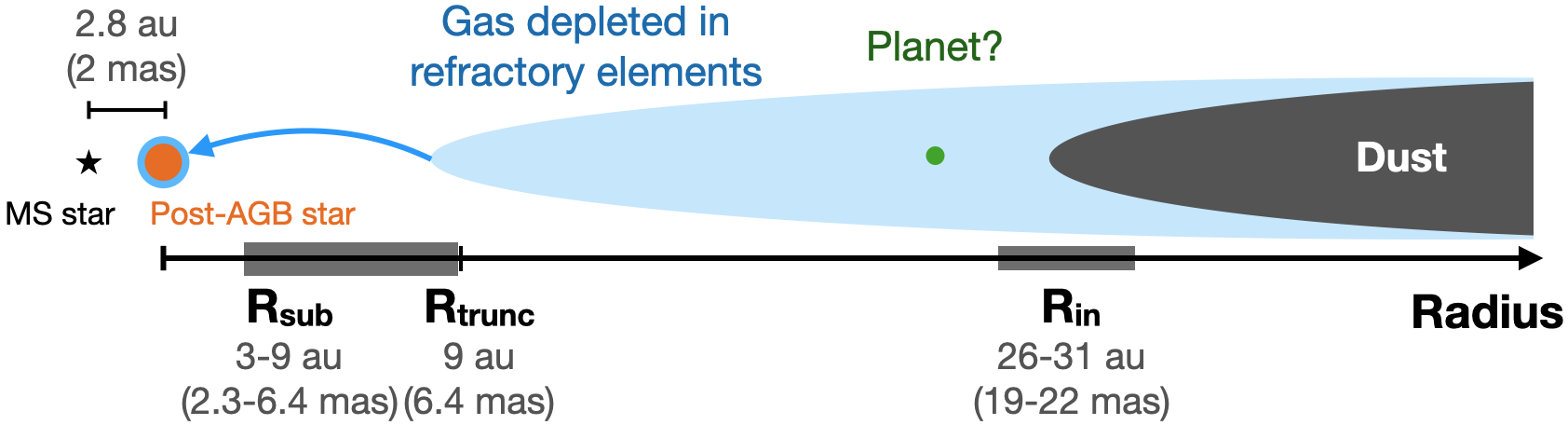}
\caption{Sketch of the AC\,Her system including the radial sizes of the different disk truncation mechanisms at work compared to the derived inner binary orbit.  This sketch  is adapted from \citep{Kluska2022}.  The orbital separation and the expected truncation and dust sublimation radii are derived in this work. The disk inner rim radius is from \citet{Hillen2015}.   Note that the angular scales are independent of luminosity, while the translation to physical sizes is distant dependent.}
\label{Figure:sketchACHer}
\end{figure*}

The disk around AC\,Her is the best-studied example of post-AGB disks with such a large  inner rim cavity $30^{+7}_{-4}$\,au  \citep{Hillen2015}. 
 This inner rim size was the result of radiative transfer modeling taking into account the angular size derived from MIDI observations \citep{Hillen2015}.

Several processes can influence the size of the inner disk cavity and in the next section, we evaluate these mechanisms.

\subsection{Dust sublimation}

First of all, the dust grains sublimate when they are too close to the star,   which creates a disk cavity that is indeed observed in the near-infrared disks around young stars \citep[e.g.,][]{ Monnier2002ApJ...579..694M,Dullemond2010ARA&A..48..205D,2017A&A...599A..85L, Lazareff2017A&A...599A..85L}.

 \citet{Kluska2019} compared  the  inner rim radius measured in H-band  wavelengths of several post-AGB objects (23) to the theoretical dust sublimation radius similarly to the analyses of young stellar objects. Their comparison of ``size-luminosity" relation reveal that for the most of the post-AGB disks, the inner rim size is indeed ruled by the dust sublimation \citep[Fig.\,4,][]{Kluska2019}. The deviated objects like AC\,Her are suspected to be transition disks.

 The theoretical sublimation radius for the AC\,Her disk was computed by previous authors with a maximum of value of 5\,au \citep{Hillen2015, Kluska2019, Kluska2022}.  Here, we estimate upper-bound of the $R_{\rm subli}$, 
from the size-luminosity trend in \citep{Kluska2019}, using conservative luminosity estimates.

  The determination of the luminosity is very dependent on the the distance estimate as well as on the total reddening and the latter is difficult to quantify for pulsating stars with a large amplitude in which part of the reddening can be circumstellar. We integrate the dereddened photosphere (E(B-V)=0.37) and by assuming the Gaia DR3 distance we obtain a $L_* = 4600L_\odot$.
This is larger than what the period-luminosity-color relation would give $L_* = 2475 L_\odot$ \citep{Bodi2019}, but rather low in comparison to the luminosity expected for a white dwarf of $0.7M_\odot$ \citep[see Fig. 11 of][]{MillerBertolami2016} which is around $10^4L_\odot$.

 For $L_*=10^4L_\odot$, we get $R_{\rm subli} = \,9\pm1$\,au by cross-matching with the ``size-luminosity" relation, Fig.\,4 of \citet{Kluska2019}. The $R_{\rm subli} = 5$\,au from \citet{Hillen2015} and the upper-limit estimation $\sim9$\,au   are significantly smaller ($\sim3$ to $\sim5$ times) than the measured lower-bound cavity radius ($26$\,au) ruling out the dust sublimation as the origin of the disk cavity.

\subsection{Dynamical truncation of the disk by the binary}

Another mechanism that might truncate the disk is the dynamical interaction with the inner binary.
The resonance torques exerted by the inner binary push the disk farther out, and the disk's viscous evolution tries to close this cavity \citep{Artymowicz1994ApJ...421..651A}.
The cavity size increases with the binary eccentricity and binary mass ratio $q=M_1/M_2$, and decreases with the disk viscosity. 
The radius size of this cavity (the inner rim radius) is estimated to be between 1 to 4 times the size of the semi-major axis of binary \citep{Artymowicz1994ApJ...421..651A, Hirsh2020MNRAS.498.2936H}. 
For AC\,Her, we found in Section\,\ref{Sec:Results_Orbit} an eccentricity of $e=0.2$ and a mass ratio $q$ of 0.5. 
From Fig.\,9 of \citet{Hirsh2020MNRAS.498.2936H}, we estimate the cavity radius to be 3.2 times larger than the binary semi-major axis.
The semi-major axis being $a\sim$2\,mas, we estimate the truncation radius $R_\mathrm{trunc}$ to be 6.4\,mas, that is $\sim$9\,au for a distance of 1.4\,kpc.
Such a radius is still $\sim3\times$ smaller than the observed lower bound cavity radius ( $\sim26$\,au). Therefore, dynamical truncation is  not sufficient either to explain the observed cavity size.

\subsection{Photoevaporation}

As discussed in \citet{Kluska2022}, photoevaporation is unlikely to create the disk cavities as there is no evidence for high energy photons in the system: the post-AGB star is too cool, with effective temperature ranged from $T_\mathrm{eff}=5225\pm125$ to $5800 \pm 250$\,K \citep{Hillen2015,VanWinckel1998,Kluska2022}. 

\begin{figure}[!htb]
\centering
\includegraphics[width=0.49\textwidth]{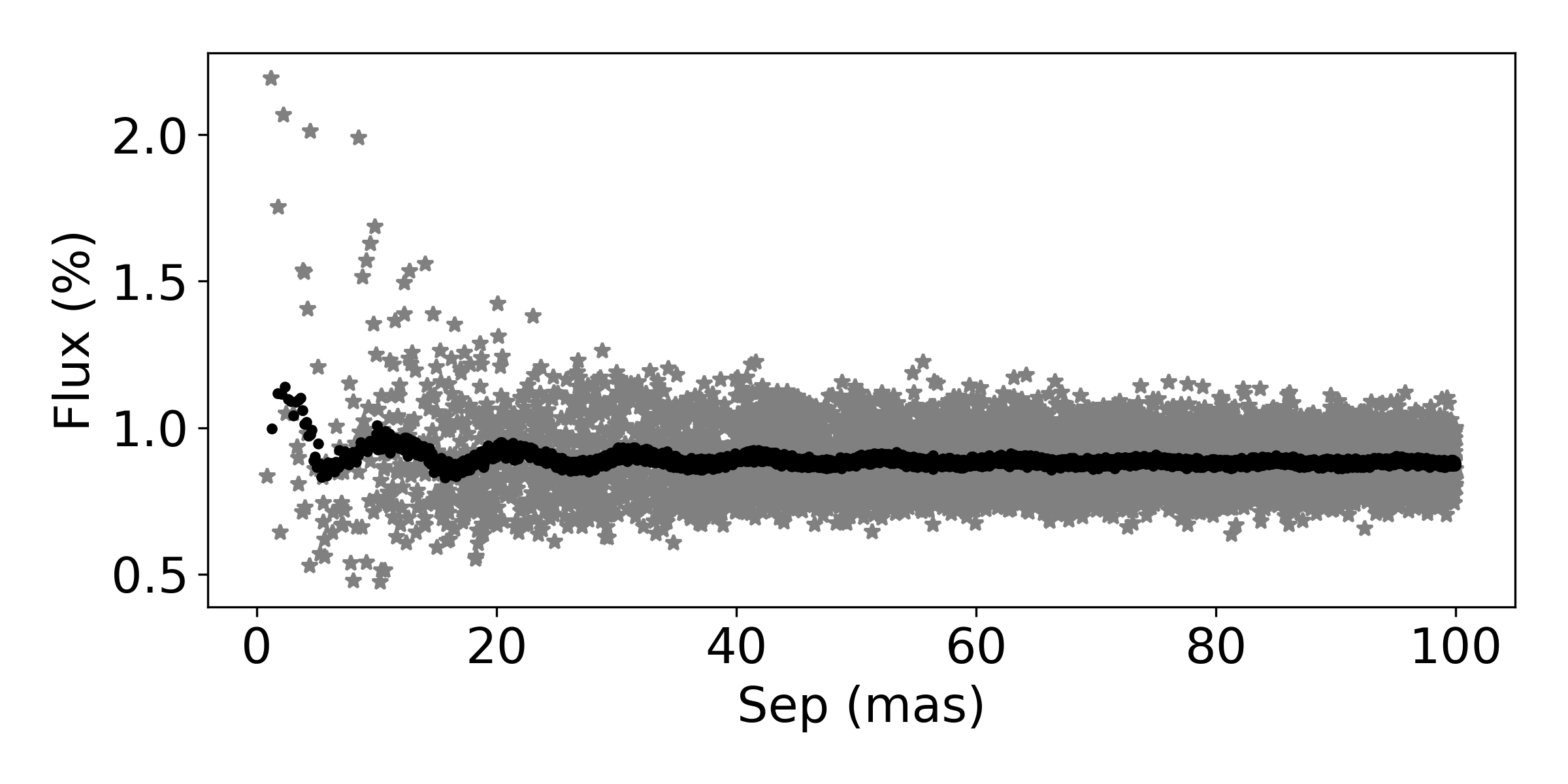}
\caption{Derived detection limits for epoch 2021 Jul 08, as a function of radius from the central star. The flux is a percentage of the primary post-AGB star.   }
\label{Figure:detection_limit}
\end{figure}

\subsection{Presence of a third (planetary) body}

Figure~\ref{Figure:sketchACHer} describes the dust sublimation and binary truncation radii in the context of the observed disk cavity and our current understanding of the systems configuration. 
The presence of (sub)planetary-mass bodies  is the most promising explanation for cavities observed in transition disks around young stars \citep{Pinilla2012A&A...545A..81P}, especially since the detections of planets in the transition disks of PDS\,70 and  
 LkCa\,15 \citep{Muller2018A&A...617L...2M,Haffert2019NatAs...3..749H,Keppler2018A&A...617A..44K,Benisty2021ApJ...916L...2B,2015Natur.527..342S} were reported.
A planet of sufficient mass can disturb the disk by exchanging angular momentum with it.
This will create a gap or cavity in the disk resulting in a pressure maximum outside the planetary orbit.
Such a pressure gradient is thought to be very efficient in trapping the dust grains while letting the gas accrete onto the star(s)   
 \citep{Pinilla2012A&A...545A..81P,Zhu2013ApJ...768..143Z, Pinilla2016A&A...596A..81P}.
It was shown that such a process is efficient in trapping the dust outside the orbit of the planet \citep[e.g.,][]{Francis2020ApJ...892..111F}.

  The surface of the post-AGB star of this system is also depleted in refractory elements, pointing towards a mechanism which traps the dust in the disk while also creating the gap \citep{Kluska2022}. We therefore conclude that a massive planet is the most likely scenario to date to explain both the cavity and the depletion of refractory elements, also in AC\,Her.

With our CHARA Array observations of AC\,Her, we are able to put an upper limit of detecting a point source inside this cavity.
We used the \texttt{PMOIRED} \texttt{detectionLimit} feature of removing the detected companion and injecting an artificial companion with random flux and position in the field of view and measure flux leading to $3\sigma$ detection \citep{Gallenne2015}. 
Figure~\ref{Figure:detection_limit} shows the detection level as a function of the position of the third component (left) and of its relative brightness so that it would have been detected at a $3\sigma$ level. 
Here, the median third star has a $3\sigma$  detection limit of 6.0\,mag. 
For AC\,Her, \texttt{PMOIRED} could have detected the tertiary if it would be brighter than the contrast difference of $\Delta H = 6$ with respect to the post-AGB star.
Such contrast puts a limit of $\sim$9$L_\odot$ on the luminosity of a putative companion that is way above the planetary   luminosity regime.
We conclude that we cannot provide a stringent constraint on the presence of a third companion in the disk cavity with these observations.

\section{Summary}
In this study, we have presented CHARA/MIRC-X observations of AC\,Her that resolve the inner binary system by exploiting sub-milliarcsecond angular resolutions in the near-infrared.

     We derived the first three-dimensional orbital solution for a post-AGB binary system. Subsequently, we constrained the dynamical system masses of the post-AGB primary ($0.73 \pm 0.13 M_\odot$) and companion ($1.40 \pm 0.12 M_\odot$).

We computed the theoretical sublimation radius $R_{\rm subli}$ using  conservative luminosity estimates and the binary truncation disk cavity radius $R_{\rm trunc}$ using the binary orbital parameters and then compared both these values with the previously measured inner disk radius.
We found that neither the dust sublimation nor the binary truncation explains the large disk cavity.   
This further strengthens the hypothesis that a planet might be clearing the disk cavity and might be responsible for the observed depletion of refractory elements on the post-AGB star.
However, our observations cannot reach enough contrast to set stringent constraints on the presence or mass of a third companion in the cavity.

    We are planning to study this system with CHARA/MYSTIC K-band observations~\citep{Monnier2018SPIE10701E..22M, Setterholm2022SPIE12183E..0BS}. 
    The K-band observations probing the dust can help further constrain the  circum-companion emission (see Section\,\ref{Sec:BinarySearch}).

    High-angular resolution observations in the thermal infrared (e.g., VLTI/MATISSE) and millimeter wavelengths (e.g., ALMA) are crucial to study the disk structure in detail and further probing the planetary hypothesis by revealing the detailed structure of the inner disk and the cavity.

We thank the anonymous referee for their constructive comments and suggestions that substantially improved the clarity of the paper.
This work is based upon observations obtained with the Georgia State University Center for High Angular Resolution Astronomy Array at Mount Wilson Observatory. The CHARA Array is supported by the National Science Foundation under Grant No. AST-1636624, AST-1715788, and AST-2034336. Institutional support has been provided by the GSU College of Arts and Sciences and the GSU Office of the Vice President for Research and Economic Development. MIRC-X received funding from the European Research Council (ERC) under the European Union's Horizon 2020 research and innovation program ("ImagePlanetFormDiscs", Grant Agreement No.\ 639889). JDM acknowledges funding for the development of MIRC- X (NASA-XRP NNX16AD43G, NSF-AST 1909165) and MYSTIC (NSF-ATI 1506540, NSF-AST 1909165). NA, SK, and CD acknowledge funding from the same ERC grant and from an STFC University of Exeter PATT travel grant (ST/S005293/1).  NA acknowledges support from the Steward Observatory Fellowship in Instrumentation and Technology Development. JK acknowledges support from FWO under the senior postdoctoral fellowship (1281121N). SK acknowledges support from an ERC Consolidator Grant ("GAIA-BIFROST", Grant Agreement No.\ 101003096) and STFC Consolidated Grant (ST/V000721/1). LS acknowledges a grant from the Marcos Moshinsky foundation.  Some of the time at the CHARA Array was granted through the NOIRLab community access program proposal id: 2020A-0274 and 2021A-0267 (PI: N.\ Anugu). This research has made use of the Jean-Marie Mariotti Center SearchCal service\footnote{available at \url{http://www.jmmc.fr/searchcal_page.htm}}. This research has made use of the Jean-Marie Mariotti Center \texttt{Aspro} service \footnote{Available at \url{http://www.jmmc.fr/aspro}}. This work has made use of data from the European Space Agency (ESA) mission {\it Gaia} (\url{https://www.cosmos.esa.int/gaia}), processed by the {\it Gaia} Data Processing and Analysis Consortium (DPAC, \url{https://www.cosmos.esa.int/web/gaia/dpac/consortium}). Funding for the DPAC has been provided by national institutions, in particular, the institutions participating in the {\it Gaia} Multilateral Agreement.

\vspace{5mm}
\facilities{CHARA Array (MIRC-X),  Gaia}
\software{CANDID \citep{Gallenne2015},
          PMOIRED \citep{Merand2022ascl.soft05001M},
          Aspro}

\pagebreak
\appendix

\section{Observations and derived astrometry}\label{Sec:Observinglog}
\begin{figure*}[!htb]
\centering
\includegraphics[width=\textwidth]{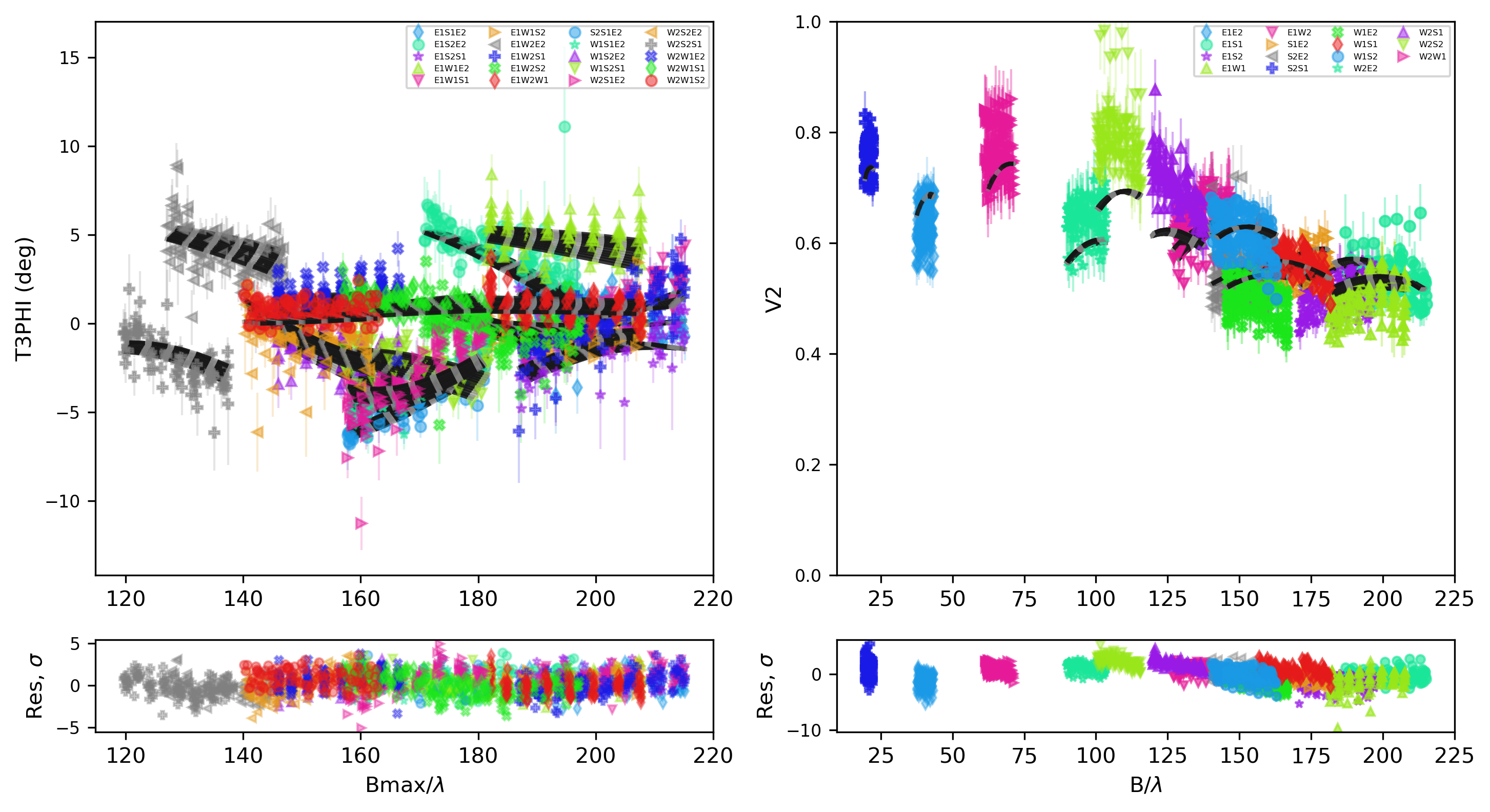}
\includegraphics[width=\textwidth]{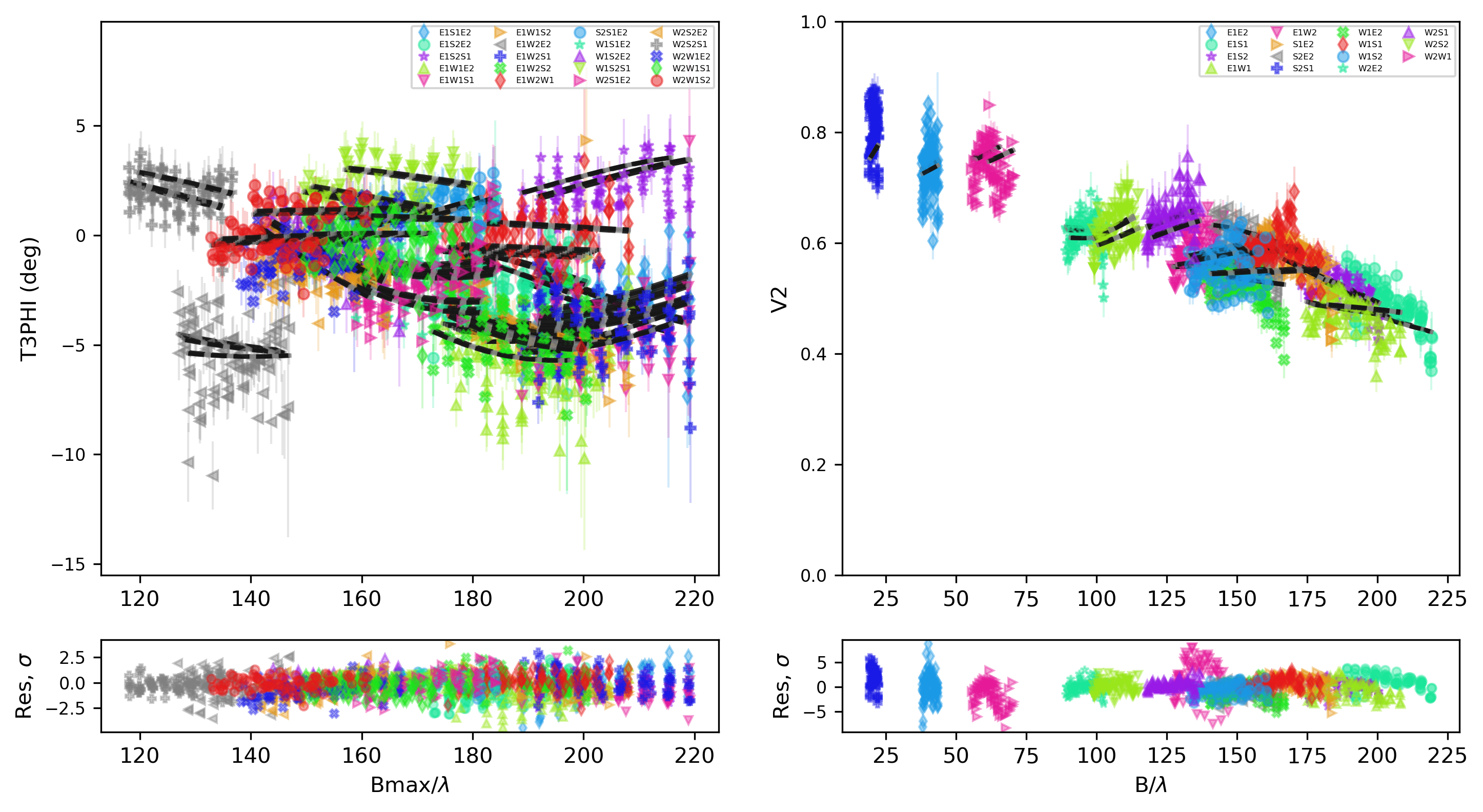}
\caption{The geometrical fit residuals to the interferometric observables recorded on AC Her with CHARA/MIRC-X  on UT 2020 Jul 31 and 2022 Jun 08.   The squared visibility and closure phase are denoted with $V^2$ and  $T3PHI$. The bottom panels are the best-fit residuals in $\sigma$. The solid black line is the model fit, and the other colors are the observed data. The bottom panels are the best-fit residuals of $T3PHI$ and $V^2$. The $V^2$ is a measure of if the object is spatially resolved, i.e.,  decreasing $V^2$ with a spatial frequency ($B/\lambda$), meaning that the observed object is spatially resolved. Non-zero closure phase indicates the source is not point symmetric, which is caused by binary and disk signals. Colors are for different baselines/closure triangles.  }
\label{Figure:fitresiduals}
\end{figure*}

Table~\ref{Table:observations} lists the CHARA/MIRC-X observing configurations and calibrators list.  The angular diameters of the calibrators are adopted from the vizier II/346/jsdc\_v2 catalog.     Figure~\ref{Figure:fitresiduals} shows the data and fitting residuals for the geometrical model described in Section~\ref{Sec:BinarySearch}. Table~\ref{Table:astrometric_positions} lists the astrometric positions, fluxes, and diameters derived from the \texttt{PMOIRED} geometrical modeling.  

\begin{figure*}[!htb]
\centering
\includegraphics[width=\textwidth]{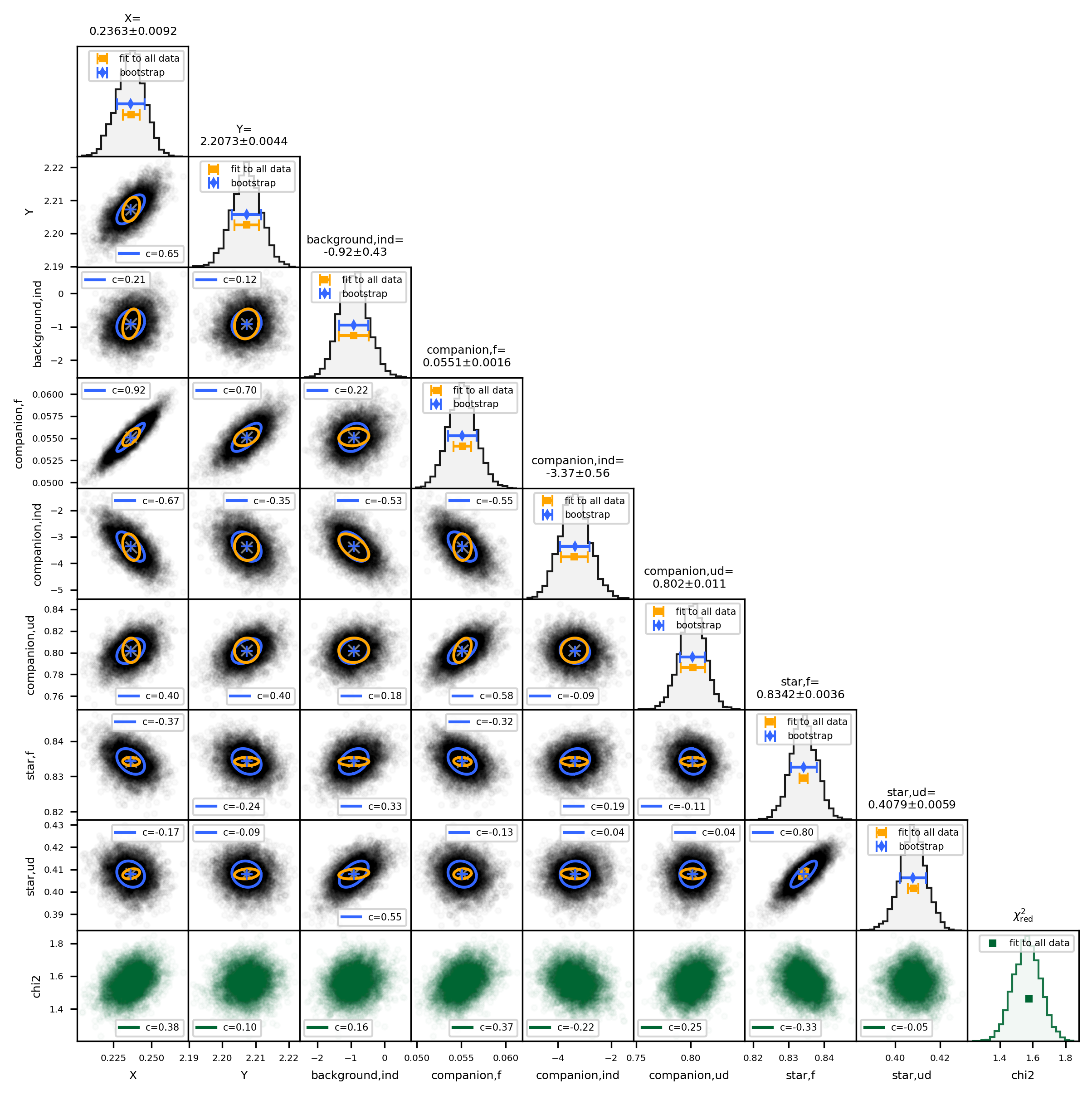}
\caption{ Estimation of uncertainties by mitigating correlated systematic using the bootstrap fitting for the 2022 Jun 08 epoch. The \texttt{"star,ud"} and \texttt{"star,f"} -- are the primary post-AGB star uniform disk diameter in mas and its flux. The \texttt{"companion,ud"},  \texttt{"companion,f"} and \texttt{"companion,ind"} -- are the companion star uniform disk diameter in mas, its flux and its spectral index. The \texttt{"X"} and \texttt{"Y"} astrometric differential positions of the companion. The \texttt{"background,ind"} is the index of the background emission. }
\label{Figure:boot_strap}
\end{figure*}

The uncertainties of the geometrical model parameters were determined using the bootstrap fitting algorithm available in \texttt{PMOIRED}. This algorithm estimates realistic uncertainties by doing a bootstrap on the data and multiple fits to estimate the scatter of the fitted parameters and mitigate the effects of correlated data. We used 5000 multiple fits. The bootstrap fits are filtered using a recursive sigma clipping algorithm. Figure~\ref{Figure:boot_strap} presents the corner plot where we used 4.5 sigma clipping.

\begin{table}
\caption{Observation log. CHARA MIRC-X observations of the AC Her binary system with observing configurations and calibrators. UT date indicates the universal time at the beginning of the data recording.  }             
\label{Table:observations}      
\centering                          
\begin{tabular}{c c c c c c}
\hline
UT Date &  Telescopes combined & Calibrators \\
\hline
2019-03-25   & 4T (No W1 \& W2) & HD~169573 \\
2019-06-07  & 6T  & HD~169573                             \\
2019-08-13   & 6T  & HD~174414, HD~181603                \\
2019-08-14   & 6T  & HD~166842, HD~178032, HD~184275      \\
2020-07-29   & 6T  & HD 174414, HD 178798                 \\
2020-07-30   & 6T  & HD 166842, HD 163506, HD 178798       \\
2020-07-31   & 6T  & HD 166842, HD 169573                 \\
2021-07-17   & 6T  & HD~161268, HD~169573, HD~166730      \\
2022-06-08    & 6T &  HD~184275, HD~194403                 \\
2022-06-17    & 6T & HD~154942, HD~167132                \\
\hline
\end{tabular}
\vskip12pt
\end{table}

\begin{table*}
\caption{The main sequence companion star astrometry (separation $\rho$ and position angle $\theta$), flux ($f_{\rm comp}$) and uniform diameter ($UD_{\rm comp}$) derived from the MIRC-X epochs. The position angle is measured East of North for the vector from the post-AGB star to the main sequence star. The primary post-AGB star is unresolved with size $UD_{\rm star}<0.5$\,mas with flux ($f_{\rm star}$). UT indicates the universal time of the observations.   }  \label{Table:astrometric_positions}           
\centering                          
\begin{tabular}{c c c c c c c c}
\hline
 UT Date & MJD & $\rho$ (mas)  & $\theta$ ($^\circ$)  & $f_{\rm comp}$ (\%) &  $UD_{\rm comp}$ (mas) & $f_{\rm star}$  \\
\hline
2019-03-25 & 58567.4765 & $2.13\pm0.10$ &     $3.7\pm3.2$ & $5.8\pm0.2$    & $0.7\pm0.2$ & $86.8\pm 0.2$  \\
2019-06-07 & 58641.3992 & $2.22\pm0.02$ &   $347.5\pm0.5$ & $5.1\pm0.2$    & $0.7\pm0.3$ & $87.5\pm 0.1$  \\
2019-08-13 & 58708.2094 & $2.12\pm0.04$ &   $335.2\pm1.0$ & $4.7 \pm 0.2$ & $0.8\pm0.3$  & $87.4\pm 0.3$ \\
2019-08-14 & 58709.2368 & $2.15\pm0.03$ &   $334.1\pm0.8$ & $4.5\pm0.2$    & $0.8\pm0.2$  & $82.1\pm 0.2$ \\
2020-07-29 & 59059.2516 & $1.46\pm0.01$ &   $191.9\pm0.8$ & $4.5\pm0.3$    & $0.8\pm0.2$  & $86.1\pm 0.1$  \\
2020-07-30 & 59060.2374 & $1.49\pm0.04$ &   $193.4\pm0.8$ & $4.5 \pm 0.1$   & $0.8\pm0.2$  & $84.6\pm 0.3$ \\
2020-07-31 & 59061.1899 & $1.45\pm0.02$ &   $193.3\pm1.0$ & $4.2\pm0.3$    & $0.7\pm0.3$  & $84.5\pm 0.1$  \\
2021-07-17 & 59412.2863 & $1.80\pm0.05$ &    $83.3\pm1.1$ & $4.1 \pm 0.2$  & $0.7\pm0.2$  & $87.8\pm 0.2$ \\
2022-06-08 & 59738.3842 & $2.22\pm0.02$ &     $6.2\pm0.5$ & $5.1\pm0.2$    & $0.8\pm0.2$  & $84.7\pm 0.1$ \\
2022-06-17 & 59747.2839 & $2.22\pm0.02$ &     $3.3\pm0.4$  &$6.0\pm0.2$   & $0.7\pm0.2$  & $84.5\pm 0.1$ \\
\hline
\end{tabular}
\vskip12pt
\end{table*}

\pagebreak
\bibliography{References}{}
\bibliographystyle{aasjournal}

\end{document}